# FX Resilience around the World: Fighting Volatile Cross-Border Capital Flows

Louisa Chen,[a*] Estelle Xue Liu[b] and Zijun Liu[c]

[a*] *Department of Accounting and Finance, University of Sussex Business School, University of Sussex, Brighton, United Kingdom. Tel: +44 (0)1273 877807. Email: l.x.chen@sussex.ac.uk;*

[b] *European Department, International Monetary Fund, Washington, D.C. 2043. Email: ELiu@imf.org*

[c] *Hong Kong Monetary Authority, Two International Finance Centre, 8 Finance Street, Hong Kong. Email: zliu@hkma.gov.hk*

# FX Resilience around the World: Fighting Volatile Cross-Border Capital Flows

**Abstract**

We show that capital flow (CF) volatility exerts an adverse effect on exchange rate (FX) volatility, regardless of whether capital controls have been put in place. However, this effect can be significantly moderated by certain macroeconomic fundamentals that reflect trade openness, foreign assets holdings, monetary policy easing, fiscal sustainability, and financial development. Passing the threshold levels of these macroeconomic fundamentals, the adverse effect of CF volatility may be negligible. We further construct an intuitive FX resilience measure, which provides an assessment of the strength of a country's exchange rates.

## 1. Introduction

Is cross-border capital flows volatility a major contributor to exchange rate volatility? How can countries subdue the effects of capital flow volatility on exchange rate volatility, while preserving the benefits that capital flows offer in terms of access to global financial resources? This is what we attempt to find out in this paper.

Cross-border capital flows have increased dramatically since 2000. While capital flows could support economic growth, the growth in capital flows was also accompanied by an increase in the volatility of these capital flows (Forbes, 2012). Extreme movements in capital flows – whether in the form of sharp increases or decreases – can lead to excessive exchange rate volatility and hence undermine economic growth and financial stability.[1]

As reviewed in the next session, existing theoretical literature generate testable implications on FX volatility determination. We explore the link between macroeconomic fundamentals and exchange rates volatility in search of moderators that can alleviate capital flow shocks passing on to exchange rate volatility. We focus on portfolio investment fund flows as they are the main contributory factor to short- and medium-term FX movement, rather than other types of capital flows such as FDI or bank flows[2,3]. Using a sample of 20 emerging market economies and advanced economics in the period from 2002:Q1 to 2020:Q3,[4] we proceed with this study in two steps. In Step 1, we conduct a cluster analysis on the impulse response of FX volatility to structural CF volatility shocks between two clusters countries in the short-term (in weeks), where countries are categorized by using the K-means clustering algorithm. We test the hypothesis that structural CF volatility shocks cause a smaller increase in FX volatility in the short-term for countries with stronger macroeconomic fundamentals, regardless of FX regimes.

---

[1] See, for example, Obstfeld and Rogoff (1998) and Gabaix and Maggiori (2015).
[2] See, for example, Brooks *et al*. (2004), Caporale *et al*. (2017), Rafi and Ramachandran (2018) and Cesa-Bianchi *et al*. (2019a,b).
[3] In the rest of the paper, capital flows refer to portfolio investment flows (equities and debts).
[4] We exclude Eurozone countries, because the fund flows data of Eurozone countries involves intra-Eurozone fund flows, making them unsuitable to our research objective.

In Step 2, we carry out a threshold analysis using panel data regressions to assess how certain macroeconomic factors could help to moderate the reaction of FX volatility to capital flow volatility in the medium-term (in quarters). We first identify economic variables that have the moderating effect, then estimate the threshold of the moderating factors[5] above which the adverse effect of CF volatility may be negligible. We further construct an FX resilience measure using the first principal component of the moderating factors.

We have three main findings: Firstly, we find that FX volatility responds to structural CF volatility shocks at a smaller scale for countries with strong macroeconomic fundamentals in the short-term, regardless of FX regimes. FX regimes only affect the components of the FX volatility response: global, rather than country-specific, capital flow shocks dominate the FX response for countries with less-strong fundamentals (mostly EMEs) under the floating FX regimes, but the is the other way round under the managed FX regimes. Secondly, by removing highly correlated economic factors[6], we identify six economic factors that can significantly dampen the adverse effect of CF volatility on FX volatility in the medium-term horizons. These moderating factors include trade openness, FX reserves, total foreign investment (the sum of net FDI, net foreign equities and net foreign bonds), (low) short-term interest rate, fiscal sustainability and financial market development. Passing certain threshold level of these moderating factors, the adverse effect of CF volatility may be negligible.

---

[5] In the rest of the paper, moderating factor refers to the economic factor that can moderate (aggravate) the adverse effect of CF volatility on FX volatility with a higher (lower) level of such a factor.
[6] This refers to credit to private sector, which is highly correlated with financial development, because credit to private sector reflects financial market depth, and financial market depth is one element of the financial development index as calculated in Svirydzenka (2016).

Finally, we find that the FX resilience measure is intuitive, in the sense that countries having high ranking of the estimated FX resilience measure coincide with those cluster countries that have smaller FX volatility reaction to the CF volatility shocks. The major contributors to the FX resilience measure are short-term interest rate, fiscal sustainability and financial market development.

The remaining part of the paper is organized as follows: Section 2 reviews the related literature. Section 3 describes the sample data and summary statistics. Sections 4 and 5 present the cluster analysis and the threshold analysis respectively. Section 6 introduces the FX resilience measure. Section 7 briefly concludes our findings and offers policy implications.

## 2. Literature review

Our paper relates to the literature on the determination of exchange rate volatility, with an emphasis on capital flow volatility.

### *2.1 Theory insights on exchange rate volatility determinants*

Our study is related to three strands of theory on FX volatility in the structural model framework. The first strand is the optimum currency area theory (OCA) originated in the seminar paper of Mundell (1961) and its later variants, who argue that shock absorption within a heterogeneous group of countries is easier if monetary and exchange rate policies remain independent, especially for countries with rigid labour markets and low international labour mobility. Variables suggested by the OCA theory, including trade interdependence and the degree of commonality in economic shocks, have considerable explanatory power for patterns of FX volatility and interventions across countries.

The predictions of the OCA theory have been acknowledged by a large number of theoretical and empirical works on exchange rate behaviour and the choice of exchange rate regime (e.g., Devereux, 2004). Devereux and Lane (2003) argue that financial linkages between countries have an important effect on desired bilateral FX volatility, beyond the standard OCA factors. Recent theoretical studies also point out that accounting for monetary and financial conditions of the domestic country helps to reduce real FX volatility differences among developing and industrial countries (Ganguly and Breuer, 2010).

The second related strand of FX volatility theory is linked to the New Open Economy Macroeconomics (the redux model) in the international finance literature, where FX volatility can be also explained by non-monetary factors such as productivity shocks, demand shocks (e.g., government spending) and labour supply shocks. The extent and composition of international trade and financial integration would smooth the impact of shocks to real exchange rate volatility (Obstfeld and Rogoff, 1995; and Hau 2000). The third related strand of FX volatility theory is the monetary approach. Dornbusch (1976) shows that M2 and interest rate changes have a stabilizing effect on the exchange rates. Later studies in a similar spirit explain exchange rates may be more volatile than underlying exchange-market fundamentals due to monetary factors (e.g., Flood, 1981).

These related theoretical works imply that economic size, trade openness, external financial linkages, government spending and internal financial development influence FX volatility. Cross-border capital flows and its volatility relate to a country's external financial linkages and dependence, no doubt play a vital role in the transmission mechanism of FX volatility.

*2.2 Empirical studies on exchange rate volatility determinants*

Empirical literature has provided evidence for the economic determinants of FX volatility suggested by the structural model in explaining long-term exchange rate movements, while not so successful in explaining short-term exchange rate movements.[7] In the last two decades, more empirical studies link capital flows to FX volatility in both developing and industrial countries, as rapid changes of capital flows were observably associated with excessive FX volatility and the consequent banking and currency crises, especially in developing countries (Kodongo and Ojah, 2012; Combes *et al.*, 2012; Ghosh *et al.*, 2016; and Caporale *et al.*, 2017). These studies find a dynamic relationship between capital flows and exchange rates, where capital flows are associated with real exchange rate appreciations and exchange rate volatility.

The impact of cross-border capital flows on the volatility of exchange rates could be driven by a range of non-fundamental factors. Economies under fixed exchange rate regimes and capital controls tend to experience lower exchange rate volatilities under capital flow shocks, as their monetary authorities are obliged to maintain the stability of the exchange rate by deploying multiple tools such as direct intervention in the foreign exchange market (e.g., Flood and Rose, 1995; and Jeanne and Rose, 2002). Investment returns and equity market volatility attract or deter speculative capital flows hence influence exchange rate movements (Grossmann *et al.*, 2014).[8] Table OA1 in the Online Appendices summarizes the related theoretical and empirical literature on the important FX volatility determinants and their relation to the FX volatility, from which our hypothesis is drawn.

---

[7] Instead, market microstructure literature explains short-term exchange rate movements in terms of market frictions and news information arrival process (e.g., Andersen and Bollerslev, 1997; Ito *et al.*, 1998; and Candian, 2019)

[8] It's worth noting that foreign exchange rates, on the other hand, could influence investors' decisions and thus cross-border capital flows (Caporale *et al.*, 2015). Since portfolio rebalancing behaviour is stronger under higher levels of exchange rate volatility, this influence could occur with some time delay (Camanho *et al.*, 2020).

### 2.3 *Contributions*

We contribute to the literature in several ways: Firstly, to the best of our knowledge, this is the first paper that quantifies the impact of CF volatility on FX volatility across the major AEs and EMEs using both high frequency (weekly) and low frequency (quarterly) data. Secondly, we identify macroeconomic fundamental factors and estimate the threshold level of these factors that can significantly moderate CF shocks on FX volatility not only in the medium-term, but also in the short-term. Finally, we introduce an FX resilience measure that is capable to assess the strength of exchange rates in the presence of volatile capital flows.

## 3. Data

### 3.1 *Sample countries and data period*

Our sample selection is mainly determined by the data of fund flows provided by Informa Financial Intelligence's Emerging Portfolio Fund Research (EPFR) database. The EPFR database contains weekly data on equity fund flows and bond fund flows of 57 countries in the period from December 1999 to September 2020, at the time of data collection. The start date and the continuity of the time series varies among the 57 countries within the original data. While half of the countries have equity fund flow data from 2002, most of the countries show bond fund flows data from 2011. To strike a balance between sample size and time span, we specify a sample period from January 2002 onwards that consists of 27 countries with full data on equity fund flows, or on bond fund flows, or on both fund flows for each country.

We further exclude Eurozone countries, because the fund flows data of Eurozone countries involves intra-Eurozone fund flows that is unsuitable for our research question – a substantial proportion of the capital flow volatility of Eurozone countries may be attributed to intra-Eurozone fund flows and yet has no influence on the Euro. The final sample consists of 20 countries for the period 01/01/2002 – 30/09/2020, including 9 Advanced Economies (AEs) and 11 Emerging Market Economies (EMEs) as categorized in the MSCI ACWI Index. See Table OA2 in the Online Appendices for the 20 sample countries.[9]

The adopted FX regimes varies across the sample countries. As a measure of a country's FX regime, we use the IMF Annual Report on Exchange Arrangements and Exchange Restrictions (AREAER), where the value of the FX regime index ranges from 1 (i.e., the least flexible FX regime) to 6 (i.e., free float FX regime). We categorize them as managed and free float FX regimes, where "managed" corresponds to an FX regime index value between 1 and 5 in the AREAER. Table 1 in the following Section 4.2 presents the sample countries' FX regimes over the sample period. For capital account openness and capital control, we use the restrictions index on capital flows (all asset categories) as in Fernández *et al*. (2016), of which the calculation is also based on the AREAER. The value of the restriction index is bounded between 0 and 1, of which a lower value refers to less restrictions in capital account and hence a lower level of capital control.

*3.2 Economic data*

The main variables in our study are capital flow (CF) volatility and foreign exchange rate (FX) volatility, where capital flow refers to the sum of equity and bond portfolio flows. As mentioned above, we use the EPFR portfolio flows (equities and bonds) as a proxy of capital flow. Following Pagliari and Hannan (2017) and Broto *et al*. (2011), we calculate capital flow volatility as the standard deviation of portfolio flows at weekly frequency (estimated in a 4-week rolling window) and quarterly frequency (estimated in a non-overlapping quarterly window). The EPFR defines portfolio flows as the percentage change in assets under management, where assets under management are subtracted by portfolio performance and foreign exchange rate change. We label this measure as $Vol_{CF}$. We calculate the volatility of nominal foreign exchange rate index in the same manner, and label this as $Vol_{FX}$.[10]

---

[9] MSCI ACWI Index is an internationally recognized benchmark for emerging and developed markets by practitioners and researchers. As of December 2019, the index covers more than 3,000 constituents across 11 sectors and approximately 85% of the free-float-adjusted market capitalization in each market.

As discussed in Section 2, in addition to capital flows, macroeconomic fundamentals and other non-fundamental factors that may affect exchange rate movement. We collect data from various sources and construct the timeseries of these variables for each country. See Tables OA3 – OA5 in the Online Appendices regarding the definitions, data sources, summary statistics and correlation matrix of these variables.

## 4. Cluster analysis

In this section, we conduct a cluster analysis to investigate if medium-term (quarterly) economic factors are related to short-term (weekly) FX volatility response to CF volatility shocks, hence be able to stabilize FX volatility in the short-term.

---

[10] Following Pagliari and Hannan (2017) and Broto *et al*. (2011), we use other two estimators to measure the volatilities of capital flows and exchange rates - the residuals of ARIMA (1,1,0) and GARCH (1,1). The average values of the ARIMA volatility estimates are much smaller than those of the standard deviation volatility estimates for both capital flows and exchange rates. To better capture the co-movement of FX volatility and CF volatility, we adopt the latter in the following analysis. Nonetheless, as a robustness test, we use the ARIMA volatility estimates to do the following cluster analysis and the threshold analysis on the composite moderating factor, of which similar conclusions are obtained. As for the GARCH (1,1) estimates, the estimation for some capital flows time series do not converge, which may be caused by the relatively sparse weekly data. We therefore do not proceed with this estimator either.

**4.1** *Approach of the cluster analysis*

We proceed with the cluster analysis in two steps. In Step 1, we normalize the quarterly fundamental economic factors (as shown in Section 3) of each country to z-scores; then we feed the normalized economic factors into the K-means clustering algorithm, allowing the algorithm to identify two distinct clusters of countries. Countries in the same cluster feature similar levels of fundamentals.[11] We expect that the fundamentals in one cluster will outperform those in the other cluster. In order to control for the direct influence of FX regime on FX volatility, we cluster countries with free float and managed FX regimes separately. In Step 2, we apply Pedroni's (2013) structural panel VARs (SPVARs) to test the hypothesis that, in the short-term (in weeks), clusters with stronger macroeconomic fundamentals have smaller FX volatility responses to the CF volatility shocks relative to the clusters with less-strong macroeconomic fundamentals, after controlling for the FX exchange regimes.

The SPVARs are fit for our data because of the cross-sectional dependencies and dynamic heterogeneities embedded in the multi-country panels. This method further allows us to decompose the impulse response into *global* and *idiosyncratic* components - the impulse responses to common capital flow shocks that capture global events (e.g., changes in global financial cycles driven by monetary policies in core countries) and to country-specific shocks that capture local events (e.g., independent monetary policies). In the rest of the paper, we designate the composite structural shock and its two components – global structural shock and idiosyncratic structural shock – as composite shocks, global shocks, and idiosyncratic shocks, respectively. The endogenous variables and their order in the SPVARs, as well as the exogenous variables are described as the following,

$$\{endogenous: Vol_{CF}, Vol_{FX}; exogenous: EquityReturn, VIX, CommodityReturn\}, \qquad (1)$$

where the endogenous variables include $Vol_{CF}$ and $Vol_{FX}$, and the exogenous variables include Equity Return, VIX and Commodity Return that are suggested in the literature as control variables (e.g., Lustig *et al*., 2011; Grossmann *et al*., 2014; and Ghosh *et al*., 2016). $Vol_{CF}$ is the standard deviation of the weekly portfolio flows (equities and bonds); $Vol_{FX}$ is the standard deviation of the weekly J.P.Morgan nominal effective exchange rate index; *EquityReturn* is the logarithmic return of domestic equity indices; *VIX* is the implied volatility of the US S&P 500 index, and *CommodityReturn* is the logarithmic return of the Bloomberg commodity index. Detailed definitions of the variables can be found in Table OA3 in the Online Appendices. We specify a 4-lag SPVARs with short-run restrictions. The short-run restrictions are based on ordering the endogenous variables in the VAR according to their speed of reaction to CF and FX volatility shocks, as FX volatility may lag respond to CF volatility shocks (e.g., Camanho *et al*., 2020).[12] We winsorize the data at the 1% and the 99% to control for outliers in the data.

---

[11] In the K-means clustering algorithm, centroids – defined as the centre of each cluster – are randomly placed; data points of the economic factors are assigned to clusters based on which centroid it is located closest to. This closeness is measured by the squared root of the sum of the distance of data points to the associated centroid. These steps are repeated until convergence, during the process the centroids are relocated to the average of the data points within each cluster. The algorithm allows the analysis and interpretation of the clusters in the context of the economic factors fed in. The Matlab code of the K-means clustering algorithm is provided by Alex Pienkowski at the IMF. This Matlab code involves a built-in K-means function of Matlab, of which the detail can be found at https://uk.mathworks.com/help/stats/K-means.html.

**4.2 *Results of the cluster analysis***

---

[12] To test the order of the endogenous variables in equation (1) and the related endogeneity issue, we conduct the Dumitrescu and Hurlin (2012) panel data Granger causality test on the winsorized $Vol_{CF}$ and $Vol_{FX}$ at one lag. The result shows that $Vol_{FX}$ does not significantly Granger cause $Vol_{CF}$, yet $Vol_{CF}$ Granger causes $Vol_{FX}$ at a high significance level. This result is consistent with the recent literature that FX volatility does not significantly affect contemporaneous equity portfolio rebalance (Camanho *et al*., 2020).

Table 1 presents the categorized clusters - the result of Step 1. As expected, Table 2 shows that one cluster (i.e., Cluster 2) outperforms the other one (i.e., Cluster 1) with higher values of the average (quarterly) economic factors over the sample period in general.

**Table 1. K-means clusters**

This table presents the result of the categorization of countries using the K-means clustering algorithm. Panels A and B refer to the two cluster countries within the free float and managed FX regime framework, respectively.

| *Panel A. Free float FX regime* | | *Panel B. Managed FX regime* | |
|---|---|---|---|
| Country | K-means Cluster (1,2) | Country | K-means Cluster (1,2) |
| Brazil | 1 | Czech Republic | 1 |
| Chile | 1 | Egypt | 1 |
| India | 1 | Malaysia | 1 |
| Thailand | 1 | Morocco | 1 |
| Australia | 2 | Russia | 1 |
| Canada | 2 | China | 1 |
| Japan | 2 | Hong Kong | 2 |
| Korea | 2 | Singapore | 2 |
| Sweden | 2 | | |
| Switzerland | 2 | | |
| United Kingdom | 2 | | |
| United States | 2 | | |

   For countries with a free float FX regime, Panel A of Table 2 shows that Cluster 2 has stronger macroeconomic fundamentals than Cluster 1, except that Cluster 2 has lower levels of real GDP growth, trade openness and FX reserves (presented in bold). Similarly, for countries with managed FX regime, Panel B shows that Cluster 2 outperforms Cluster 1 in terms of all economic factors except for real GDP growth. Panel C shows that AEs by and large have stronger macroeconomic fundamentals than EMEs, except for real GDP growth and FX reserves. Based on these results, we designate Cluster 2 and AEs as the clusters with strong macroeconomic fundamentals, and Cluster 1 and EMEs as the clusters with less-strong macroeconomic fundamentals.

**Table 2. The mean value of the economic factors of the K-means clusters, and EMEs *vs* AEs.**

This table presents the median values of the quarterly economic factors across countries within each cluster. Numbers in bold are *inconsistent* with the hypotheses of strong macroeconomic fundamentals. All variables are defined in Table OA3 in the Online Appendices.

| Economic factors, panel median, 2002Q1-2019Q4 | RealGDPGrowth | TradeOpenness | FXReserves | TFI | CreditPrivate | ShortRate | FiscalSurplus | FinanicalDevelopment | FXRegime | |
|---|---|---|---|---|---|---|---|---|---|---|
| High/Low as to strong macroeconomic fundamentals | High | High | High | High | High | Low | High | High | - | Conclusion |
| *Panel A. Countries with free float FX regime* | | | | | | | | | | |
| Cluster 1 | 1.0 | 11.4 | 0.3 | 0.3 | 86.5 | 5.0 | -0.6 | 0.5 | 6 | Cluster 2 has stronger macroeconomic fundamentals than Cluster 1, given the same flexibility of the FX regimes. |
| Cluster 2 | **0.6** | **10.7** | **0.1** | 1.3 | 174.6 | 1.1 | -0.3 | 0.9 | 6 | |
| *Panel B. Countries with managed FX regime* | | | | | | | | | | |
| Cluster 1 | 1.3 | 12.4 | 0.3 | 0.5 | 83.0 | 3.0 | -0.8 | 0.5 | 3 | Cluster 2 has stronger macroeconomic fundamentals than Cluster 1, given the similar flexibility of the FX regimes. |
| Cluster 2 | **0.9** | 74.0 | 1.2 | 4.7 | 165.1 | 0.5 | 0.8 | 0.7 | 3 | |
| *Panel C. EMEs vs. AEs* | | | | | | | | | | |
| EMEs | 1.1 | 12.8 | 0.3 | 0.5 | 89.1 | 3.2 | -0.7 | 0.5 | 5 | AEs has stronger macroeconomic fundamentals and more flexible FX regimes than EMEs. |
| AEs | **0.6** | 13.3 | **0.1** | 1.7 | 170.2 | 0.7 | -0.2 | 0.9 | 6 | |

Figure 1, as the result of Step 2, presents the impulse response of FX volatility to the structural CF volatility shocks estimated by using equation (1),[13] where the composite CF volatility shocks (row 1) is further decomposed into the global shocks (row 2) and idiosyncratic shocks (row 3). For countries adopting free float FX regimes, Figure (A.1) shows that Cluster 1 countries (featured with less-strong fundamentals) has relatively higher and bumpy FX volatility responses to the composite CF volatility shocks than Cluster 2 countries (featured with strong fundamentals). This

---

[13] All the five variables do not contain unit roots up to lag 5 in the Levin-Lin-Chu (2002) panel unit root tests.

is more pronounced in the first three weeks after the shocks. The difference in the response is attributed to the component of global shocks (Figure A.2) more than the component of idiosyncratic shocks (Figure A.3), implying that, under a floating FX regime, countries with less-strong fundamentals (mostly EMEs) are more sensitive to the global shocks than those with strong fundamentals (mostly AEs).

This pattern also holds for countries with managed FX regimes (Panel B) and for the group of AEs and EMEs (Panel C).[14] The only difference is that the FX volatility response is mainly driven by the component of idiosyncratic shocks (Figures B.3 and C.3) rather than the component of global shocks (Figures B.2 and C.2). This implies that, managed FX regimes can more or less shelter countries which have weak fundamentals from changes arising from global financial cycles. Meanwhile, those countries become more vulnerable to their domestic shocks.

**Figure 1. The weekly impulse response function of FX volatility to capital flow volatility shocks**

This figure shows the weekly impulse response function of FX volatility to the structural shocks of capital flow volatility estimated by using equation (1). Panels A and B refer to the two cluster countries within the free float and managed FX regime framework, respectively. Panel C refer to EMEs vs AEs. The solid line depicts the median impulse responses among the sample of countries. The lower and upper edges of the shaded area represent the $25^{th}$ and $75^{th}$ percentile responses among the sample of countries.

---

[14] The cumulative impulse response of FX volatility to the CF shocks between the two comparative clusters is more sizeable. See Figure OA1 in the Online Appendices.

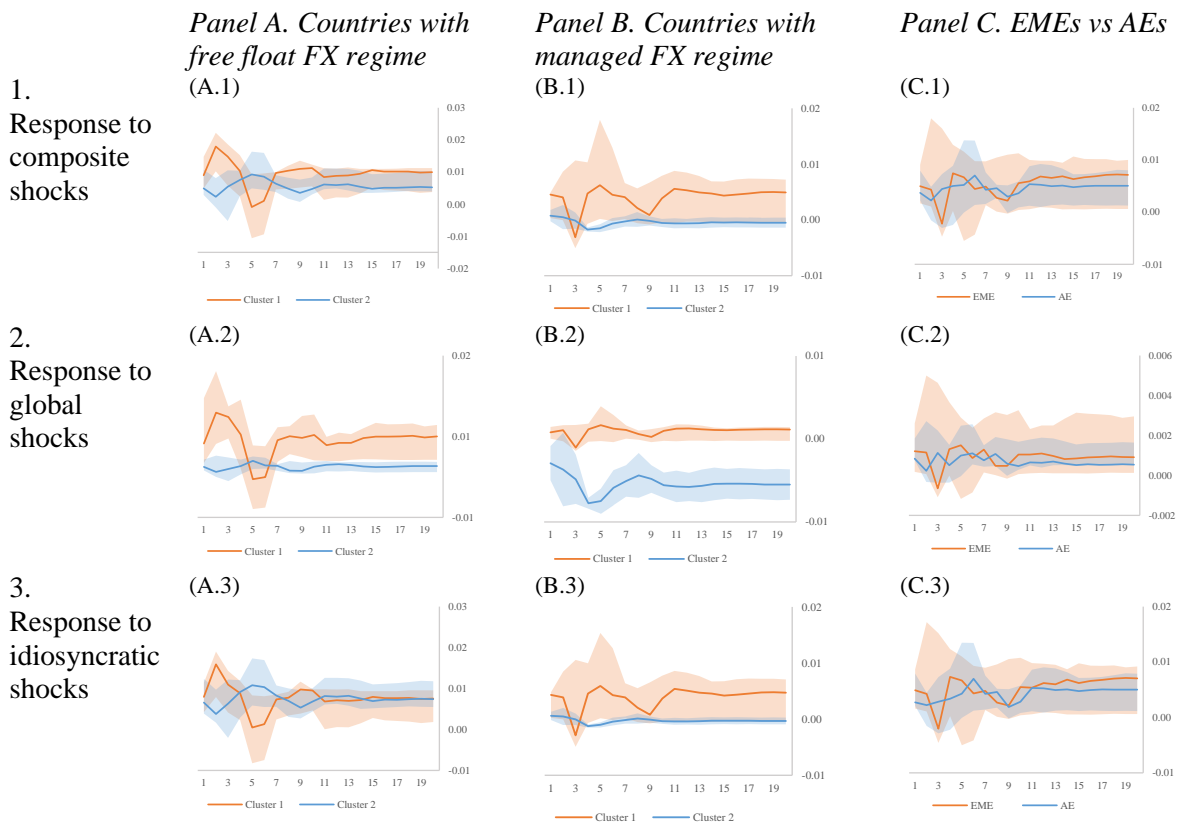

Overall, the cluster analysis supports our hypothesis that countries with stronger macroeconomic fundamentals are more capable of absorbing CF volatility shocks hence stabilizing FX volatility in the short-term. Our finding supports the theoretical argument as discussed in McDonald (1999), that macroeconomic fundamentals play an important role in explaining exchange rate movements even in short-term horizons.[15]

## 5. Threshold analysis

---

[15] We use Uhlig's (2005) Bayesian approach to estimate equation (1) with sign restrictions. The result shows consistency with those of the Pedroni's panel SVAR (see Section 7 in the Online Appendices).

In this section, we use panel data regression to test the moderating effect of the economic factors on FX volatility in the presence of CF volatility over medium-term horizons (quarterly). We first lay out the procedure of the threshold analysis, followed by results and discussions.

## *5.1 Procedure of the threshold analysis*

As discussed in the previous sections, some fundamentals not only reflect a country's economic advances, but also link to the movement of its exchange rate. We first investigate how these factors influence FX volatility in a constant CF volatility environment through the following panel data regression,

$$Vol_{FX_{it}} = \beta_0 + \beta_1 Vol_{CF_{it}} + \boldsymbol{\beta_2 MF_{it}} + \boldsymbol{\beta_3 X_{it}} + \mu_t + \eta_i + \varepsilon_{it}, \qquad (2)$$

where $\boldsymbol{MF_{it}}$ is a vector of the moderating factors (i.e., economic variables) used in the cluster analysis in the previous section, including real GDP growth, trade openness, FX reserves, total financial investment (the sum of net FDI, net portfolio equities and bonds), short-term interest rate, fiscal surplus and financial development. We exclude credit to private sector because of its high correlation with financial development; $\boldsymbol{X_{it}}$ is a vector of control variables, including FX regime index, capital control (i.e., capital account openness), local equity return, capital flow and VIX; $\mu_t$ is the time-specific effect; $\eta_i$ is country-specific effect and $\varepsilon_{it}$ is the error term. We use feasible generalized least squares estimators to adjust for the presence of AR(1) autocorrelation within panels, as well as cross-sectional correlation and heteroskedasticity across panels.

Secondly, since the main focus is to investigate the potential cushioning effect of the economic variables on CF-volatility-induced FX volatility, we expand equation (2) to the following equation (3) by including an interaction term of *Vol_CF* and its multiplication with one of the seven economic variables in turn,

$$Vol_{FX_{it}} = \gamma_0 + \gamma_1 Vol_{CF_{it}} + \gamma_2 Vol_{CF_{it}} * MF_{it} + \boldsymbol{\gamma_3}(\boldsymbol{other\ MF_{it}}, \boldsymbol{X_{it}}) + \mu_t + \eta_i + \varepsilon_{it}, \quad (3)$$

where $Vol_{CF_{it}} * MF_{it}$ is the interaction term; ***other MF*** $_{it}$ is a vector of the moderating factors excluding the one involved in the interaction; other variables are defined in equation (2). In this equation, the total effect of $Vol_{CF}$ on $Vol_{FX}$ is given by $(\gamma_1 + \gamma_2 MF_{it})$, and the total effect of a specific MF on $Vol_{FX}$ is given by $(\gamma_2 Vol_{CF_{it}} + \boldsymbol{\gamma_3})$. We interest in the former, as it tells us that, holding all other variables constant (insulating the marginal effect of other variables), the effect of $Vol_{CF}$ on $Vol_{FX}$ depends on not only its elasticity to $Vol_{FX}$, but also the level of the specific MF in the interaction term. For the specific *MF* (e.g., FX reserves), we expect $\gamma_2 < 0$, so that the total effect of CF volatility - $(\gamma_1 + \gamma_2 MF_{it})$ - becomes more negative when CF volatility increases. *Vice versa*, for the aggravating *MF* (i.e., short-term interest rate), we expect $\gamma_2 > 0$, so that the total effect CF volatility becomes more positive when CF volatility increases. This means that a higher (lower) level of the mitigating (aggravating) MF could exert the moderating effect of CF volatility on FX volatility.

Thirdly, to consider the simultaneous moderating effect of all the MFs, we construct a composite MF by using the first principal component of the significant MFs through the principal component analysis. The method is given below,

$$PC1MF_{i,t} = \sum_{m=1}^{6}(\boldsymbol{\omega_m MF_{i,m,t}}), \quad (4)$$

where ***MF*** is a vector of the significant moderating factors normalized between [0, 1], $\boldsymbol{\omega_m}$ is the estimated loading of the first principal component of the respective *MF*, *i* is the country identifier, and *t* is the year-quarter time identifier. We estimate the threshold of the composite MF associated with CF volatility as following:

$$Vol_{FX_{it}} = \gamma_0 + \gamma_1 Vol_{CF_{it}} + \gamma_2 Vol_{CF_{it}} * PC1MF_{it} + \gamma_3 PCIMF_{it} + \mu_t + \eta_i + \varepsilon_{it}, \quad (5)$$

where all variables are defined in equations (3) and (4).

Finally, we estimate the threshold of the economic factors above which that the adverse effect of CF volatility can be reduced to a low level, even be negligible. Following the approach in Aghion *et al.* (2009), when the estimated $\gamma_1$ and $\gamma_2$ are statistically significant, we can estimate the threshold of *MF* (i.e., $\widetilde{MF}$) above which the total effect of *Vol*$_{CF}$ on *Vol*$_{FX}$ is no more than a target level, $\theta$:

$$\partial(Vol_{FX_{it}})/\partial(Vol_{CF_{it}}) = \gamma_1 + \gamma_2 MF_{it} < \theta \Leftrightarrow MF_{it} > \widetilde{MF} := (\theta - \gamma_1)/\gamma_2 \qquad (6)$$

For example, if setting the target level $\theta = 0$, a 1% increase in *Vol*$_{CF}$ will have nearly no effect on *Vol*$_{FX}$, given the respective MF (i.e., the MF in the interaction term) is above the threshold $-\gamma_1/\gamma_2$, regardless of other MFs. Similarly, if setting the target level $\theta < 0.05$, a 1% increase in *Vol*$_{CF}$ will cause no more than 0.05 increase in *Vol*$_{FX}$, given the respective MF is above the threshold $(0.05 - \gamma_1)/\gamma_2$, regardless of other MFs.

## *5.2 Results of the threshold analysis*

Regression (1) of Table 5 shows the estimates of equation (2). Except short-term interest rate and total foreign investment have significant impact on *Vol*$_{FX}$, most of the economic factors have no significant influence on FX volatility in a constant CF volatility environment.

However, this picture changes in a CF heightened environment. As shown in the upper part of regressions (2-8) in Table 5, the interaction term ($\gamma_2$) is highly significant for all MFs, except real GDP growth. This implies that, most of the economic factors are capable to moderate the CF-volatility-induced FX volatility. The Wald test shows that the total effect of *Vol*$_{CF}$ is different from zero at a high significance level across all MFs, confirming the joint significance of $\gamma_1$ and $\gamma_2$. We discuss in more details below.

Regressions (3-5) and (7-8) of Table 5 show that trade openness, FX reserves, total foreign investment, fiscal surplus and financial development are the moderating variables (MFs) involved in the interaction term respectively, where $Vol_{CF}$ is the predictor variable. All these MFs obtain highly significant and negative $\gamma_2$ of the interaction term, meaning one unit increase in the respective MFs can reduce the adverse effect of (a given) $Vol_{CF}$ on $Vol_{FX}$ by -0.002, -0.02, -0.02, -0.07 and -0.21, respectively. Turning to the aggravating factor, namely short-term interest rate, regression (6) of Table (5) shows that the coefficient of the interaction term, $\gamma_2$, is 0.01 and highly significant, implying that a 1% lower of the short-term interest rate can reduce the adverse effect of (a given) $Vol_{CF}$ on $Vol_{FX}$ by 0.01. Regressions (3-8) of Figure 2 below illustrate that the total effects of $Vol_{CF}$ on $Vol_{FX}$ become more negative when the level of these MFs becomes higher (lower for short-term interest rate).

Most of these moderating effects are economically important. Firstly, a well-developed financial market offers more tools for investors to hedge against exchange rate risks, thus dampening currency instability. Secondly, a sustainable fiscal position provides governments with more room for monetary policy adjustment, and hence more scope to respond to adverse capital flow shocks. During an investment boom funded by capital inflows, fiscal contraction reduces domestic demand and cools down capital inflows. Financial markets therefore reward more sustainable fiscal positions. The IMF found that declines in inflows between 2010 and 2015 were 1 percentage point of GDP larger for countries with above-average public debt ratios.[16] Thirdly, high foreign investment (FDI, portfolio equities and bonds) and FX reserves are traditional standby

---

[16] See Chapter 2, IMF April 2016 World Economic Outlook.

## Table 3. Estimates of the FX volatility moderating factors and their thresholds

In this table, regression (1) presents the estimates of equation (2); regressions (2 – 8) present the estimates of equations (3), in which the interaction term is the multiplication of the specific moderating factor and $Vol_{CF}$; regression (9) presents the estimates of equation (5). The thresholds are estimated by using equation (3), where both coefficients of $Vol_{CF}$ and the interaction term are required to be, at least, at 10% significance level. $PCIMF$ is defined in equation (4). Threshold is defined in equation (6). Other variables are defined in Table OA3 in the Online Appendices. ***, ** and * refer to the 1%, 5% and 10% significance level, respectively.

| Dependent variable: $Vol_{FX}$ | (1) | (2) | (3) | (4) | (5) | (6) | (7) | (8) | (9) |
|---|---|---|---|---|---|---|---|---|---|
| | | MF | | | | | | | |
| | | RealGDPGrowth | TradeOpenness | FXReserve | TotalForeignInvestment | ShortRate | FiscalSurplus | FinanicalDevelopment | PC1MF |
| $Vol_{CF}$ | 0.02*** | 0.02* | 0.06*** | 0.03*** | 0.08*** | -0.04*** | 0.02* | 0.15*** | 0.22*** |
| $Vol_{CF}$*MF | | 0.00 | -0.002*** | -0.02*** | -0.02*** | 0.01*** | -0.07*** | -0.21*** | -0.19*** |
| RealGDPGrowth | -0.01 | -0.01 | -0.01* | -0.01 | -0.02 | -0.01* | -0.01 | -0.01 | |
| TradeOpenness | 0.00 | 0.00 | 0.00 | 0.00 | 0.00 | 0.00 | 0.00 | 0.00 | |
| FXReserve | 0.00 | 0.00 | 0.00 | 0.01 | 0.00 | 0.00 | 0.00 | -0.01 | |
| TotalForeignInvestment | -0.01* | -0.01* | -0.01** | -0.01* | 0.01 | -0.01* | -0.01** | -0.01** | |
| ShortRate | 0.04*** | 0.04*** | 0.04*** | 0.04*** | 0.04*** | 0.02* | 0.04*** | 0.04*** | |
| FiscalSurplus | -0.03 | -0.03 | -0.03 | -0.03 | -0.05 | -0.03 | 0.06*** | -0.01 | |
| FinanicalDevelopment | -0.3 | -0.31 | -0.11 | -0.2 | 0.31 | 0.08 | 0.07 | 0.15 | |
| PC1MF | | | | | | | | | -0.27* |
| *Combined effects: Wald tests (p-values)* | | | | | | | | | |
| $H_0$: $Vol_{CF}$ total effect = 0 | - | 0.02 | 0.00 | 0.10 | 0.00 | 0.01 | 0.00 | 0.00 | 0.00 |
| $H_0$: **MF** total effect = 0 | - | 0.25 | 0.87 | 0.52 | 0.58 | 0.01 | 0.47 | 0.88 | 0.00 |
| *Threshold analysis: Moderating effect of MF on VolFX* | | | | | | | | | |
| a) When marginal $Vol_{CF}$ effect on $Vol_{FX}$ smaller than zero ($\Theta = 0$) | | | | | | | | | |
| **MF threshold** greater than | - | - | 52.0 | 5.8 | 13.0 | 1.5 | 0.5 | 0.9 | 1.3 |
| *s.e.* | - | - | *0.01* | *0.01* | *0.07* | *0.01* | *0.01* | *0.01* | *0.01* |
| b) When marginal VolCF effect on VolFX smaller than 0.05 ($\Theta < 0.05$) | | | | | | | | | |
| **MF threshold** greater than | - | - | 6.0 | -1.2 | 2.0 | 8.5 | -0.5 | 0.5 | 0.9 |
| *s.e.* | - | - | *0.01* | *0.03* | *0.02* | *0.01* | *0.01* | *0.02* | *0.01* |
| Sample mean | | | 20.0 | 0.5 | 1.7 | 3.4 | -0.5 | 0.7 | 1.04 |

**Figure 2. The total effect of CF volatility on FX volatility**

This Figure is associated with Regressions (3-8) of Table 5, where the total effect of CF volatility on FX volatility (the solid line) is plot against the associated moderating factor (MF). The 90% confidence intervals are the shaded areas.

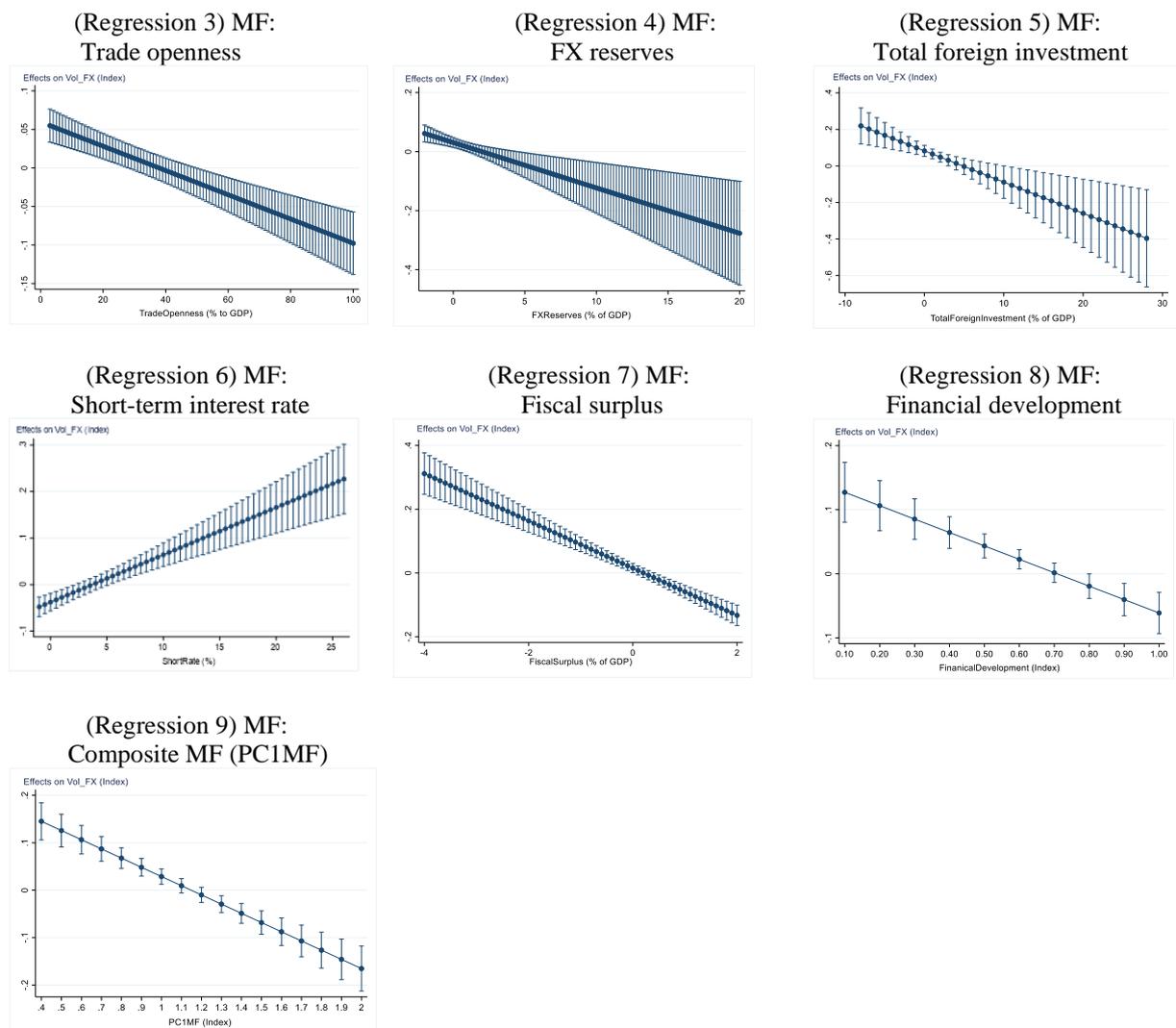

tools acting as counter measures for large exchange rate movements. Their moderating effect persists in an environment with heightened capital flow volatility.

Finally, interest rate is a key determinant of FX volatility from monetary approach viewpoints. The level of interest rates is intimately connected with international capital flows in and out countries with high or low interest rates, altering the supply and demand of the currencies in the markets, thus affecting FX volatility (Flood, 1981). The positive relationship between short-term interest rates and FX volatility has two implications. On the one hand, lower interest rates are associated with loose monetary and low inflation environment, which strengthens firms' profitability and an optimistic macroeconomic forecast. The upbeat economic outlook reduces uncertainty and lowers exchange rate volatility. On the other hand, low interest rates imply that interest rate differentials between domestic and anchored country markets become smaller, that deters speculative short-term capital inflows and smooths out capital flows in both directions. Grossmann *et al*. (2014) and Alper *et al*. (2013) provide similar arguments for advanced and emerging markets.

The lower part of regressions (2-8) of Table 3 reports the estimates of equation (6) - the threshold of individual MF at $\theta = 0$ and $\theta < 0.05$. The standard errors of the respective thresholds are computed using a delta method, namely by taking a first-order Taylor approximation around the mean. The lower part of Regressions (3) shows that, to moderate the total effect ($\theta$) of $Vol_{CF}$ on $Vol_{FX}$ towards 0, trade openness would need to be higher than 38%, holding other economic variables constant. This threshold is much higher than the sample mean of 20%, as we insulate the influence from other MFs. Similar findings are held for FX reserves, total foreign investment and fiscal surplus whose thresholds are 2%, 5% and 0.2%, respectively. The threshold of financial development is an exception, with a value of 0.8 close to the sample mean of 0.7, implying that most of the countries have acknowledged the importance of financial market development and have advanced the improvement.

The lower part of Regression (6) in Table 3 shows that, by setting the total effect of $Vol_{CF}$ on $Vol_{FX}$ ($\theta$) towards 0, the short-term interest rate should be lower than 3.5%, assuming other economic variables are unchanged. In other words, if a country's short term interest rate is lower than 3.5%, a 1% increase in CF volatility will not significantly push up FX volatility, holding other economic variables constant. Note that all the estimated thresholds obtain small standard errors of no more than 0.03. Assuming the effect ($\theta$) of $Vol_{CF}$ on $Vol_{FX}$ is more tolerant towards 0.05, the associated thresholds of the MFs become lower (see the lower part of Table 3).

Our findings suggest that the moderating effect on exchange rate volatility merges from the trade and the financial channels. The latter is more profound in terms of the magnitude of the moderating effect (i.e., the coefficients of the interaction term $\gamma_2$). This concurs with the recent literature suggesting that financial factors appear to play an important role in explaining exchange rates (see Meese and Rogoff, 1983 and Cesa-Bianchi *et al*., 2019 a, b).

Regression (9) of Table 3 reports the thresholds of the composite MF as calculated by equation (4). Not surprisingly, the estimated coefficients of $Vol_{CF}$, PC1MF and their interaction term ($\gamma_2$) are highly significant, where $\gamma_2$ is -0.19. Regression (9) of Figure 2 illustrates this moderating effect, where the adverse effect of $Vol_{CF}$ decreases with an increase in $PC1MF$. The lower part of Regression (9) in Table 3 shows that, for the total effect ($\theta$) of $Vol_{CF}$ on $Vol_{FX}$ towards zero, the threshold of PC1MF would be 1.1. Similarly, for a total effect ($\theta$) of $Vol_{CF}$ on $Vol_{FX}$ towards 0.05, the threshold of PC1MF would be 0.9. The results suggest that the composite MF is an effective aggregate moderating factor.

## 6. FX resilience measure

The composite MF can be further used to construct an FX resilience measure in the context of volatile capital flows:

$$FX_{Resillience_{i,\tau}} = \partial(Vol_{FX_{i,t}})/\partial(Vol_{CF_{i,t}}) = \widehat{\gamma_1} + \widehat{\gamma_2}\,\overline{PC1MF_{i,\tau}}, \tag{7}$$

where $\overline{PC1MF_{i,\tau}}$ is the average composite moderating factor of country *i* over the period of *τ*; $\widehat{\gamma_1}$ and $\widehat{\gamma_2}$ are the estimates in equation (6); $FX_{Resillience_{i,\tau}}$ is the FX resilience measure, equal to the total effect of *Vol*$_{CF}$ on *Vol*$_{FX}$ for a given $\overline{PC1MF_{i,\tau}}$. A more negative value of $FX_{Resillience}$ indicates a smaller effect of *Vol*$_{CF}$ and a stronger FX resilience.

We estimate the FX resilience measure for the 20 sample countries over the whole sample period, where each country's $\overline{PC1MF_{i,\tau}}$ is the median of its quarterly PC1MF from 2002 to 2019. We rank the estimated FX resilience measure of the 20 countries from the lowest (the most resilient) to the highest (the least resilient). Table 4 reports the ranked measures, the associated p-value and the 90% confidence intervals.

**Table 4. FX resilience measure ranking**

This table presents the FX resilience measure and their ranking of the sample countries. Panels A1 and A2 refer to the two cluster countries within the free float and managed FX regime framework, respectively. Panel B1 and B2 refer to EMEs and AEs, respectively. FX resilience measure is estimated by using equation (7), where the composite moderating factor is the country's quarterly average over the sample period of 2002 - 2019. Capital control is defined in Table OA3 in the Online Appendices. The shaded areas refer to countries with strong macroeconomic fundamentals as defined in Tables 1 and 2. The FX risillience measures are ranked from the lowest to the highest, of which a lower value refers to a more resilient foreign exchange rate to capital flow volatility.

| Country: FX resilience measure ranking | Vol$_{CF}$ marginal effect on Vol$_{FX}$ at the median PC1MF 2002-2019 | 90% cofidence intervals | Capital Control Index | Country: FX resilience measure ranking | Vol$_{CF}$ marginal effect on Vol$_{FX}$ at the median PC1MF 2002-2019 | 90% cofidence intervals | Capital Control Index |
|---|---|---|---|---|---|---|---|
| *Panel A1. Free float FX regime* | | | | *Panel A2. Managed FX regime* | | | |
| Switzerland | -0.03 | [-0.04, -0.01] | 0.3 | Hong Kong | -0.10 | [-0.13, -0.07] | 0.1 |
| Korea | 0.00 | [-0.02, 0.01] | 0.2 | Singapore | -0.07 | [-0.10, -0.05] | 0.1 |
| Sweden | 0.00 | [-0.02, 0.01] | 0.1 | Malaysia | 0.02 | [0.00, 0.03] | 0.8 |
| Canada | 0.00 | [-0.02, 0.01] | 0.1 | Czech Republic | 0.03 | [0.01, 0.04] | 0.3 |
| Thailand | 0.01 | [-0.01, 0.02] | 0.8 | China | 0.04 | [0.02, 0.06] | 1.0 |
| Australia | 0.01 | [0.00, 0.03] | 0.3 | Russia | 0.05 | [0.03, 0.06] | 0.5 |
| United Kingdom | 0.02 | [0.00, 0.04] | 0.0 | Morocco | 0.07 | [0.04, 0.09] | 0.8 |
| Japan | 0.03 | [0.01, 0.04] | 0.0 | Egypt | 0.12 | [0.08, 0.15] | 0.2 |
| United States | 0.03 | [0.01, 0.05] | 0.2 | | | | |
| Chile | 0.04 | [0.02, 0.05] | 0.2 | | | | |
| Brazil | 0.08 | [0.06, 0.10] | 0.6 | | | | |
| India | 0.09 | [0.06, 0.11] | 1.0 | | | | |
| *Panel B1. EMEs* | | | | *Panel B2. AEs* | | | |
| Korea | 0.00 | [-0.02, 0.01] | 0.2 | Hong Kong | -0.10 | [-0.13, -0.07] | 0.1 |
| Thailand | 0.01 | [-0.01, 0.02] | 0.8 | Singapore | -0.07 | [-0.10, -0.05] | 0.1 |
| Malaysia | 0.02 | [0.00, 0.03] | 0.8 | Switzerland | -0.03 | [-0.04, -0.01] | 0.3 |
| Czech Republic | 0.03 | [0.01, 0.04] | 0.3 | Sweden | 0.00 | [-0.02, 0.01] | 0.1 |
| Chile | 0.04 | [0.02, 0.05] | 0.2 | Canada | 0.00 | [-0.02, 0.01] | 0.1 |
| China | 0.04 | [0.02, 0.06] | 1.0 | Australia | 0.01 | [0.00, 0.03] | 0.3 |
| Russia | 0.05 | [0.03, 0.06] | 0.5 | United Kingdom | 0.02 | [0.00, 0.04] | 0.0 |
| Morocco | 0.07 | [0.04, 0.09] | 0.8 | Japan | 0.03 | [0.01, 0.04] | 0.0 |
| Brazil | 0.08 | [0.06, 0.10] | 0.6 | United States | 0.03 | [0.01, 0.05] | 0.2 |
| India | 0.09 | [0.06, 0.11] | 1.0 | | | | |
| Egypt | 0.12 | [0.08, 0.15] | 0.2 | | | | |

Panel A1 of Table 4 shows the ranked FX resilience measures for countries with free float FX regimes, where Chile, Brazil and India have the lowest rankings and Thailand is in the middle of the ranking. The relatively high ranking of Thailand can be largely attributed to the country's high level of capital control, which at 0.8 is close to the maximum capital control measure of 1. Relating the FX resilience measure ranking with Tables 1 and 2, these four countries demonstrate less-strong macroeconomic fundamentals (Cluster 1). The countries featured with strong macroeconomic fundamentals (Cluster 2, shaded grey) have higher FX resilience measure ranking than aforementioned four countries. The same conclusion applies for countries with managed FX regimes and for EMEs and AEs as shown in Panels A2, B1 and B2 of Table 4. The results of the threshold analysis and cluster analysis could be systematized into one coherent body of FX resilience determinants, that some economic factors can effectively moderate capital flow shocks on exchange rate volatility, without heavily relying on rigid countermeasures such as exchange rate management and capital controls.

We evaluate the contribution of individual MFs to the FX resilience measure by their weighted first principal component. Figure 3 illustrates the contributions of the six MFs to the above estimated FX resilience measures. Panel A shows that the value of the FX resilience measure decreases with the composite MF, and Panel B plots the MFs' percentage share of contribution to the composite MF. Visually, the top 3 contributing MFs are short-term interest rate, fiscal surplus and financial development for almost all the countries, except that trade openness has bigger share for Hong Kong and Singapore due to their special status as international trade and financial centres.

The decomposition of the FX resilience measure suggests that policy makers, wishing to tackle the CF-volatility-induced FX volatility, could set their long-term priorities: maintaining relatively loose monetary policy (with credible inflation targeting regimes), improving fiscal sustainability, and advancing towards broader financial market depth and efficiency in the long term.

In summary, the results of the threshold analysis support the theoretical argument that macroeconomic fundamentals play a crucial role in stabilizing exchange rate volatility over medium-term horizons (e.g., the OCA theory; and McDonald, 1999), especially during periods of rising capital flow volatility.

**Figure 3. FX resilience measure and the contributing moderating factors**

This figure is associated with Table 4. Panel A shows the FX resilience measure and the six MFs' contributions weighted by their first principal component. Panel B shows the PC1-weighted MF contributions to the FX resilience measure in a 100% scale.

*Panel A. FX resilience measure vs PC1-weighted MF contributions for individual countries*

*Panel B. PC1-weighted MF contributions for individual countries (in 100%)*

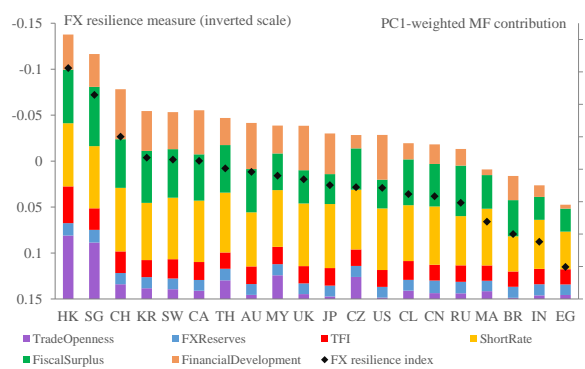
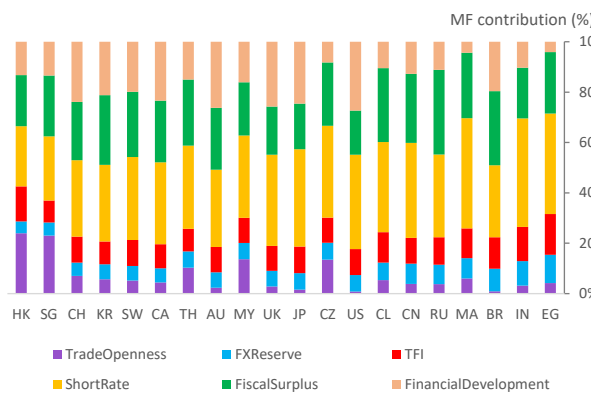

We apply the moderating factors threshold and the FX resilience measure to the recent global events – the emerging market currency crisis in 2018 and the 2020 Covid-19 pandemic – when global financial markets experienced large capital outflows and inflows. We find that the prediction based on the moderating factors threshold and the FX resilience measure are consistent with the outcomes of the two market events. For details, see Sections 8 and 9 in the Online Appendices.

## 7. Conclusion and policy recommendations

Since the 1990s, the rapid growth of cross-border capital flows has become an important feature of the global economic landscape. Capital flows have the potential to increase global welfare by allowing resources to be allocated efficiently and enabling countries to better share risk. Meanwhile, the volatility of the flows and the resulting heightened exchange rate volatilities can undermine the desired benefits (Carney, 2019).

Using a sample of 20 EMEs and AEs (excluding Eurozone countries due to the unsuitability of intra-Eurozone fund flows data), we find that certain macroeconomic fundamentals, either taken as individual factors or as an aggregate composite factor, can effectively moderate the impact of capital flow volatility on exchange rate volatility in the short- and medium-term, regardless of FX management and capital control. These economic factors are suggested by the theoretical literature on FX volatility determinants, including trade openness, FX reserves, total foreign investment (excluding FX reserves), short-term interest rate, fiscal sustainability and financial development. We further estimate the thresholds of the individual economic moderating factors and the aggregate composite moderating factor, above which the total effect of CF volatility could be negligible. By using the aggregate composite moderating factor, we establish an FX resilience measure for individual countries. The estimated FX resilience measures for the sample countries over the period from 2002 to 2019 point to three major contributing economic factors to the FX resilience measure: short-term interest rate, fiscal surplus and financial development.

Our study offers policy recommendations. Firstly, to strengthen macroeconomic fundamentals towards the threshold level could be one of the priorities to smoothen short- and medium-term exchange rate volatility in an environment of rising cross-border flows volatility. Among them, maintaining a loose monetary policy to keep low costs of capital, reducing fiscal deficit and speeding up financial market development are the most effective measures. Secondly, one imminent challenge amid the ongoing Covid-19 pandemic is how each country should prepare themselves for a sudden monetary tightening of the advanced economies, given the prospect of a global multi-speed recovery (IMF, 2021) and higher-than-expected inflation resulting from the substantial fiscal stimulus and a sharp recovery in consumption. From this perspective, our finding recommends a useful tool for policy makers to assess their exchange rate resilience associated with volatile capital flows.

While it is of pivotal importance to mitigate capital flows risk, a key challenge for policy makers is to preserve the benefits that cross-border capital flows offer to circulate financial resources in the global market so as to improve economic growth and welfare. Our study tackles this issue by offering tentative solutions to balance these objectives. As an early attempt, our paper can serve as a springboard for future studies. One possible extension would be to probe the economic factors that may help to solve or reduce the classic policy trilemma.

**Acknowledgements**

We thank David Cook, Dirk Krueger, Ranko Jelic, Robert Macrae and the participants at participants of the 8th Annual Conference of the International Association for Applied Econometrics (IAAE), the 2021 conferences on Financial Globalization and De-Globalization" in Hong Kong, the 2021 World Finance & Banking Symposium in Riga (virtual), research seminars at the International Monetary Fund and University of Sussex Business School for their comments. We also thank Peter Pedroni for sharing the SPVARs codes, as well as Harry Parker and Boyang Sun for excellent assistance on data collection. The views and opinions expressed in this article are those of the authors and do not necessarily reflect the official policy or position of International Monetary Fund or Hong Kong Monetary Authority.

# FX Resilience around the World: Fighting Volatile Cross-Border Capital Flows

Louisa Chen, Estelle Xue Liu and Zijun Liu

## Online Appendices

**Contents**

# 1.

# Table OA1. Related literature on the determinants of exchange rate volatility

| Macroeconomic fundamentals | Related economic factors in the paper | Symbols of the economic factors | Strong(+)/ Weak(-) fundamental at a high value | Positive(+)/ Negative(-) influence on FX volatility | Related literature | |
|---|---|---|---|---|---|---|
| | | | | | Theorectical | Empirical |
| Output | Real GDP growth | RealGDPGrowth | + | - | Mundell (1961)'s OCA theory and its variants | Bayoumi and Eichengreen (1998); Devereux and Lane (2003); Canales-Kriljenko and Habermeier (2004); Sutherland (2005); Glick and Hutchison (2005); Hviding et al. (2004); Hausmann et al. (2006); Grossmann et al. (2014); |
| Trade openness | Total import and export | TradeOpenness | + | - | Mundell (1961)'s OCA theory and its variants; Obstfeld and Rogoff (1995) redux model and its variants | Bayoumi and Eichengreen (1998); Devereux and Lane (2003); Canales-Kriljenko and Habermeier (2004); Hviding et al. (2004); Sutherland (2005); Hausmann et al. (2006); Cho and Doblas-Madrid (2014); Calderón and Kubota (2018); |
| Foreign assets holdings | FX reserves | FXReserves | + | - | Mundell (1961)'s OCA theory and its variants | Devereux and Lane (2003); Canales-Kriljenko and Habermeier (2004); Glick and Hutchison (2005); Melvin et al. (2009); Dominguez et al. (2012); Grossmann et al. (2014); |
| | Total foreign investment (the sum of net FDI, net portfolio equities and net portfolio bonds) | TFI | + | - | | |
| Monetary policy stance | Short-term interest rate | ShortRate | - | + | Dornbusch (1976) model and its variants; Mundell (1961)'s OCA theory and its variants (e.g., Devereux 2004). | Bhansali (2007); Giannellis and Papadopoulos (2011); Ganguly and Breuer (2010); Grossmann et al. (2014); Yung (2017); |
| Leverage | Total credit to private sector | CreditPrivate | + | - | Mundell (1961)'s OCA theory and its variants | Ganguly and Breuer (2010); Calderón and Kubota (2018) |
| Fiscal sustainability | Fiscal surplus | FiscalSurplus | + | - | The monetary approach (e.g., MacDonald and Talor, 1993) | Canales-Kriljenko and Habermeier (2004); Miyamoto et al. (2019); IMF (2021) |
| Financial development | Financial development index (including markets | FinancialDevelopment | + | - | Mundell (1961)'s OCA theory and its variants | Devereux and Lane (2003); Calderón and Kubota (2018) |
| Exchange rate regime | Flexibility of the exchange rate regime | FXRigime | N/A | - | Mundell (1961)'s OCA theory and its variants | Canales-Kriljenko and Habermeier (2004); Kocenda and Valachy (2006); Klein and Shambaugh (2008); |
| Capital control | Openness of capital account | CapitalControl | N/A | - | Obstfeld and Rogoff (1995) redux model and its variants; Reinhart (2000) | Edison and Reinhart (2001); Reinhart and Smith (2002); Melvin et al. (2009); Edwards and Rigobon (2009); Calderón and Kubota (2018) |

**Additional References**

**2.**

**Table OA2. Sample countries: AEs and EMEs**

This table lists the sample countries in our analysis, including nine advanced economies (AEs) and eleven emerging market economies (EMEs). Managed FX period is the year that the country implemented a managed FX regime. Managed FX regimes have the FX regime index below 6 and free float FX regimes have the FX regime index equals 6 as in the IMF Annual Report on Exchange Arrangements and Exchange Restrictions (AREAER).

| AEs | | | EMEs | | |
| --- | --- | --- | --- | --- | --- |
| Country | Has managed FX periods? (Yes/No) | Managed FX periods | Country | Has managed FX periods? (Yes/No) | Managed FX periods |
| Australia | No | | Brazil | No | |
| Canada | No | | Egypt | Yes | 2002-2019 excl. 2016 |
| Japan | No | | India | No | |
| Sweden | No | | Morocco | Yes | 2002-2019 |
| United Kingdom | No | | Russia | Yes | 2002-2013 |
| United States | No | | Chile | No | |
| Hong Kong | Yes | 2002-2019 | China | Yes | 2002-2019 |
| Singapore | Yes | 2002-2019 | Czech Republic | Yes | 2002-2006; 2013-2016 |
| Switzerland | No | | Korea | No | |
| | | | Malaysia | Yes | 2002-2007; 2009-2015 |
| | | | Thailand | No | |

## 3.
## Table OA3. Variables: definitions and data sources

| Variables | Variables Symbols | Definitions and measures | Unit | Used of data | Original data frequency | Data sources |
|---|---|---|---|---|---|---|
| FX volatility | $Vol_{FX}$ | Vol_FX is the S.D. of nominal effective exchange rate at weekly frequency (estimated in a 4-week rolling window) and quarterly frequency (estimated in a non-overlap quarterly window). | Index | J.P.Morgan nominal effective exchange rate index. The index measures a currency's nominal exchange rate relative to a basket of other currencies using an trade-weighted calculation. | Daily | Bloomberg and authors' calculation |
| Capital flow | CapitalFlow | Funds flow (equities and bonds) is the % change of asset under management subtracted by portforlio performance and foreign exchange rate change. | % | Equity Funds, ETFs & Mutual Funds: Portfolio Change (%); Equity Funds,ETFs & Mutual Funds: Ending Assets (EOP,Mil.US$); Bond Funds: ETFs & Mut Funds: Port Change(AVG, %) and Bond Funds: ETFs/Mut Funds: End Assets(EOP, Mil.US$). Weekly | Weekly | EPFR and authors' calculation |
| Captital flow volatility | $Vol_{CF}$ | Vol_CF is the S.D. of funds flow (equity and bond) at weekly frequency (estimated in a 4-week rolling window) and quarterly frequency (estimated in a non-overlap quarterly window). | % | The same as above. | Weekly | EPFR and authors' calculation |
| Real GDP growth | RealGDPGrowth | Real GDP growth rate calcuated as quarterly % change. | % | Gross Domestic Product Based On Purchasing Power Parity, Standardized, Constant Prices, Seasonally Adjusted | Quarterly | Datastream and authors' calculation |
| GDP | GDP | Annual GDP | Millions, USD | GDP, PPP (current international $) | Annually | World Bank - WDI |
| Net export | NetExport | Net export of goods and services | % of annual GDP | Current Account, Goods and Services, Net, Millions USD | Quarterly | IFS and authors' calculation |
| Trade openness | TradeOpenness | Total export and import divided by annual GDP | % of annual GDP | Export and Import, Current Account, Goods and Services, Credit, US Dollars | Quarterly | calculation |
| FX reserves | FXReserves | FX reserves | % of annual GDP | Supplementary Items, Reserves and Related items, US Dollars | Quarterly | IMF - IFS and authors' calculation |
| Net FDI | NetFDI | Net FDI | % of annual GDP | Financial Account, Net Lending (+) / Net Borrowing (-) (Balance from Financial Account), Direct Investment, Net Acquisition of Financial Assets, US Dollars | Quarterly | IMF - IFS and authors' calculation |
| Net portfolio equity | NetEquity | Net equities | % of annual GDP | Financial Account, Portfolio Investment, Net Acquisition of Financial Assets, Equity and Investment Fund Shares, US Dollars | Quarterly | IMF - IFS and authors' calculation |
| Net portfolio bond | NetBond | Net bonds | % of annual GDP | Financial Account, Portfolio Investment, Net Acquisition of Financial Assets, Debt Securities, US Dollars | Quarterly | IMF - IFS and authors' calculation |
| Total foreign investment | TFI | The summary of net FDI, net equtities and net bonds | % of annual GDP | See the definitions of Net FDI, Net portfolio equityes and Net portfolio bonds as described above. | Quarterly | IMF - IFS and authors calculation |
| Credit to private | CreditPrivate | Total credit to private sector | % of annual GDP | Total credit to the private non-financial sector (core debt) | Quarterly | BIS |
| Fiscial surplus | FiscialSurplus | Net government borrowing, lending(-)/borrowing(+). | % of annual GDP | General government net lending/borrowing, percent of fiscal year GDP (Percent of annual GDP for quarterly data) | Quarterly | IMF |
| Short-term interest rate | ShortRate | 3-month money market rate or Tbill rate | % | Australia, Singapore and U.S. use 3-month money market rate (code I60B.. in Datastream, has over 95% corelation with 3-month Tbill rate). The rest of the sample counties use 3-month Tbill rate from Bloomberg. | Quarterly | Datastream and Bloomberg |
| Finanical development index | FinanicalDevelopment | Finanical development index developed by IMF. See https://data.imf.org/?sk=F8032E80-B36C-43B1-AC26-493C5B1CD33B | Index | A measure of the development of finanical markets and institutions of a country in terms of their depth (size and liquidity), access (ability of individuals and companies to access finanical services) and efficiency (ability of institutions to provide financial services at low cost and with sustainable revenues and the level of activity of capital markets) | Annually | IMF |
| FX regime | FXRegime | IMF classification of FX regimes from No separate legal tender (1) to Free floating (6). | Index (1 -6) | The IMF's Annual Report on Exchange Arrangements and Exchange Restrictions (AREAER) | Annually | IMF |
| Capital control | CapitalControl | An overall restrictions index (all asset categories) on capital flows as in Fernández *et al*. (2016). | Index (0 -1) | The IMF's Annual Report on Exchange Arrangements and Exchange Restrictions (AREAER) | Annually | IMF |
| Equity market index return | EquityReturn | The logrithm return of domestic equity index, | % | MSCI index | Daily | Datastream |
| U.S. equtity volatility index | VIX | CBOE VIX index | Index | CBOE VIX index | Daily | WRDS CRSP |
| Global commodity return | CommodityReturn | The logrithm return of Bloomberg Commodity Index | Index | Bloomberg Commodity Index (BCOM) is calculated on an excess return basis and reflects commodity futures price movements. The index rebalances annually weighted 2/3 by trading volume and 1/3 by world production and weight-caps are applied at the commodity, sector and group level for diversification. | Daily | Bloomberg |

# 4.
# Table OA4. Variables: summary statistics

Variables are defined in Table OA3 in the Online Appendices.

|  | Obs | Mean | Std. Dev. | Min | Max |
|---|---|---|---|---|---|
| **2002 - 2020: Weekly** | | | | | |
| Vol_FX (Index) | 19580 | 0.56 | 0.50 | 0.04 | 17.28 |
| Vol_CF (%) | 19580 | 0.86 | 1.63 | 0.00 | 37.30 |
| EquityReturn (%) | 19580 | 0.10 | 3.50 | -48.31 | 37.74 |
| VIX (Index) | 19580 | 19.39 | 9.04 | 9.34 | 72.03 |
| CommodityReturn (%) | 19580 | -0.02 | 2.29 | -11.11 | 10.66 |
| **2002Q1 - 2019Q4: Quaterly** | | | | | |
| Vol_FX (Index) | 1440 | 1.13 | 0.98 | 0.08 | 17.86 |
| Vol_CF (%) | 1440 | 1.10 | 1.60 | 0.00 | 20.38 |
| RealGDP_Growth (%) | 1440 | 0.91 | 1.51 | -9.01 | 19.21 |
| TradeOpenness (% of annual GDP) | 1440 | 20.01 | 20.03 | 3.52 | 100.88 |
| FXReserve (% of annual GDP) | 1440 | 0.53 | 1.67 | -7.52 | 20.42 |
| TFI (% of annual GDP) | 1440 | 1.72 | 3.20 | -7.29 | 28.83 |
| CreditPrivate (% of annual GDP) | 1440 | 132.93 | 60.97 | 23.00 | 392.40 |
| ShortRate (%) | 1440 | 3.39 | 3.83 | -0.84 | 26.24 |
| FiscalSurplus (% of annual GDP) | 1440 | -0.52 | 1.05 | -3.62 | 1.99 |
| FinancialDevelopment (Index) | 1440 | 0.66 | 0.20 | 0.27 | 1 |
| FX_Regime (Index) | 1440 | 4.95 | 1.40 | 2 | 6 |
| CapitalControl (Index) | 1440 | 0.36 | 0.33 | 0 | 1 |
| CapitalFlow (%) | 1440 | 0.09 | 0.84 | -4.57 | 8.03 |
| EquityReturn (%) | 1440 | 1.62 | 12.14 | -72.02 | 46.58 |
| VIX (Index) | 1440 | 18.90 | 7.97 | 10.30 | 58.32 |
| CommodityReturn (%) | 1440 | -0.13 | 9.16 | -35.84 | 15.35 |

# 5.
# Table OA5. Variables: correlation matrix

Variables are defined in Table OA3 in the Online Appendices.

| 2002Q1 - 2019Q4 Quaterly | Vol_FX | Vol_CF | RealGDP_Growth | TradeOpenness | FXReserve | TFI | CreditPrivate | ShortRate | FiscalSurplus | FinancialDevelopment | FX_Regime | CapitalControl | CapitalFlow | EquityReturn | VIX |
|---|---|---|---|---|---|---|---|---|---|---|---|---|---|---|---|
| Vol_CF | 0.14 | | | | | | | | | | | | | | |
| RealGDP_Growth | -0.16 | 0.04 | | | | | | | | | | | | | |
| TradeOpenness | -0.24 | 0.07 | 0.08 | | | | | | | | | | | | |
| FXReserve | -0.08 | 0.02 | 0.15 | 0.26 | | | | | | | | | | | |
| TFI | -0.14 | 0.02 | 0.02 | 0.60 | 0.15 | | | | | | | | | | |
| CreditPrivate | -0.05 | -0.15 | -0.13 | 0.27 | 0.04 | 0.38 | | | | | | | | | |
| ShortRate | 0.23 | 0.26 | 0.01 | -0.31 | -0.13 | -0.21 | -0.61 | | | | | | | | |
| FiscalSurplus | -0.13 | 0.06 | 0.03 | 0.48 | 0.19 | 0.34 | 0.30 | -0.29 | | | | | | | |
| FinancialDevelopment | 0.03 | -0.19 | -0.17 | 0.10 | 0.02 | 0.23 | **0.80** | -0.50 | 0.23 | | | | | | |
| FX_Regime | 0.21 | -0.12 | -0.17 | -0.50 | -0.13 | -0.30 | 0.15 | -0.03 | -0.14 | 0.42 | | | | | |
| CapitalControl | -0.09 | 0.06 | 0.25 | -0.18 | 0.05 | -0.30 | -0.47 | 0.22 | -0.13 | -0.55 | -0.28 | | | | |
| CapitalFlow | 0.12 | 0.29 | 0.11 | 0.01 | 0.08 | 0.06 | -0.02 | 0.04 | 0.02 | -0.01 | -0.01 | 0.01 | | | |
| EquityReturn | -0.24 | -0.03 | 0.21 | -0.01 | 0.10 | 0.07 | -0.06 | -0.01 | -0.01 | -0.04 | 0.00 | 0.02 | 0.22 | | |
| VIX | 0.25 | 0.04 | -0.16 | -0.01 | -0.04 | -0.05 | -0.05 | 0.02 | -0.10 | -0.01 | 0.00 | -0.01 | -0.08 | -0.31 | |
| CommodityReturn | -0.18 | 0.00 | 0.18 | -0.01 | 0.09 | 0.06 | -0.04 | -0.02 | 0.00 | -0.03 | 0.00 | 0.01 | 0.14 | 0.51 | -0.27 |

6.

**Figure OA1. The weekly cumulative impulse response function of FX volatility to the composite capital flow volatility shocks**

This figure shows the weekly cumulative impulse response function of FX volatility to the composite structural shocks of capital flow volatility estimated by using equation (1). Figures (A), (B) and (C) are associated with (A.1), (B.1) and (C.1) of Figure 1.

(A) Countries with free float FX regimes
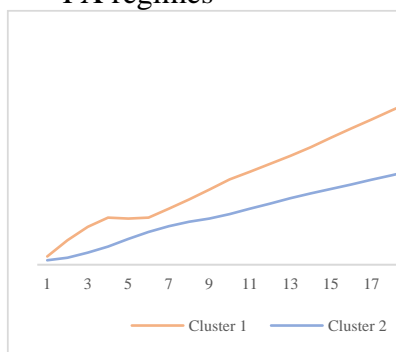

(B) Countries with managed FX regimes
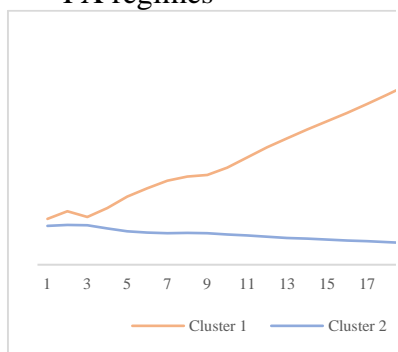

(C) EMEs *vs* AEs
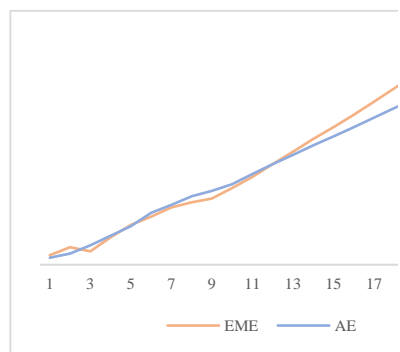

7.

**Robustness test of SPVARs**

To demonstrate the robustness of the SPVARs result to a more flexible identification scheme of structural VAR estimation, we use Uhlig's (2005) Bayesian approach to estimate equation (1) with sign restrictions. It is reasonable to assume that a shock to CF volatility causes a positive response of CF volatility and an undefined (unrestricted) response to FX volatility, while a shock to FX volatility causes both CF volatility and FX volatility to respond positively. The sign restrictions of the impulse response variables to the shocks of CF volatility and FX volatility are specified in Table OA6 below.

**Table OA6. Sign restrictions for the structural shocks of CF volatility and FX volatility**

This table presents the sign restrictions for the impulse response function of CF volatility ($Vol_{CF}$) and FX volatility ($Vol_{FX}$) to the structural shocks of CF volatility and FX volatility used in equation (1). The variables are defined in Table OA3 in the Online Appendices. "+" and "Any" refer to positive and unrestricted sign of the impulse response function, respectively.

| Shock\Sign of the response | CF volatility | FX volatility |
|---|---|---|
| $Vol_{CF_t}$ | + | Any |
| $Vol_{FX_t}$ | + | + |

Using the sign restrictions, we estimate the impulse response function for individual countries, then aggregate the CF-volatility-shock-size-weighted impulse response function of FX volatility across countries within each cluster. The results of the impulse response function of FX volatility are presented in Figure OA2 below, which shows consistency with those of the Pedroni's panel SVAR.

**Figure OA2. The impulse response function of FX volatility to capital flow volatility shocks**

This figure presents the impulse response function of FX volatility to CF volatility shocks estimated using equation (1) and Uhlig (2005) sign-restricted VAR. Figures (a) and (b) refer to the two cluster countries within the free float FX regime framework and the managed FX regime framework, respectively. The two clusters (Cluster 1 vs Cluster 2) are defined in Table 1. Figure (c) refers to the two types of countries - EMEs vs AEs. The solid line depicts the CF-volatility-shock-size-weighted impulse response function of FX volatility across constituent countries within the cluster, where the lower and upper edges of the shaded area represent the associated ±1 std CI of the responses.

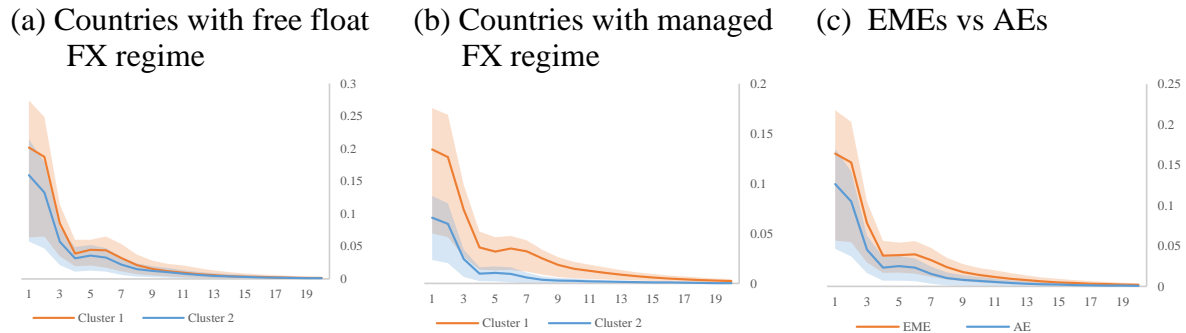

(a) Countries with free float FX regime  (b) Countries with managed FX regime  (c) EMEs vs AEs

## 8.

## Case study: The 2018 emerging market currency crisis

We use a recent global market event for a robustness test on the thresholds of the moderating factors identified and estimated in Section 5.

In 2018, amid the US monetary policy tightening and rising crude oil price, many emerging markets around the world experienced plummeting currencies and capital outflows. For instance, the Indian rupee witnessed high volatility, falling by nearly 14% between April to October in 2018. The Brazilian *real* depreciated by nearly 20% from January to September in 2018. Other countries hit by sharp currency depreciation include Argentina, Turkey, South Africa, and Indonesia. Meanwhile, some countries like Korea, Malaysia, Hong Kong, and Switzerland, weathered the storm well and maintained relatively stable exchange rates. As described below, such differences in FX volatility can be largely explained by the number of MFs that have reached to their thresholds at $\theta \leq 0$.

We proceed with this case study in two steps. Firstly, we produce a market-based FX resilience ranking for the sample countries. This ranking is based on a country's percentage change in FX volatility per unit of CF volatility between 2017 and 2018. A smaller (more negative) percentage change in FX volatility per unit of CF volatility refers to a higher market-based FX resilience ranking. Secondly, we calculate the country's median quarterly MFs in 2018 and count the number of MFs that have attained their estimated thresholds at $\theta \leq 0$ (see the values of the thresholds in Table 3) – we label them as supporting factors. Thirdly, we compare the market-based FX resilience ranking with the number of supporting factors and expect a positive relationship between the two. Table OA7 below presents the results.

Panel A of Table OA7 shows that countries with a higher market-based FX resilience ranking featured with smaller (more negative) FX volatility change (per unit of CF volatility) in 2018 and larger number of supporting factors. Countries such as Chile, Brazil and Morocco, whose currencies were severely affected by heightened FX volatility, failed to meet all the thresholds of the six MFs. On the contrary, countries had relatively lower FX volatility changes in 2018 had more supporting factors in general. To give a straightforward illustration, Panel B of Table OA7 plots the number of supporting factors against the market-based FX resilience measure for the sample countries. The downwards-sloping regression line has a negative coefficient at 1% significance level, implying that a more resilient exchange rate can be accomplished by attaining more supporting factors. Note that in this case study, we exclude Egypt who experienced FX regime change in 2018, as well as China, Malaysia, and India whose capital control was equal or greater than 0.9 (the maximum value of capital control is 1) in the same year.

**Table OA7. Case study of moderating factor thresholds: the U.S. Federal Reserve interest rate hike in 2018**

*Panel A. FX volatility moderating factors vs annual percentage change of FX volatility to per unit of CF volatility in 2018*

This table presents the change in FX volatility per unit of CF volatility between 2018 and 2017 (change in FX volatility, hereafter) *vs* the average quarterly moderating factors in 2018 of the sample countries (excluding China, Malaysia and India whose capital control is equal or greater than 0.9, and excluding Egypt who experienced FX regime change in 2018). Countries are ranked based on the change in FX volatility from the lowest to the highest. Values in bold refer to the moderating factors attaining the thresholds (i.e., supporting factors). The total number of supporting factors of each country is reported in the far-right column. MF thresholds are presented in Table 3. Variables are defined in Table OA3 in the Online Appendices.

| Country code | MF threshold when $Vol_{CF}$ marginal effect on $Vol_{FX}$ is significantly $< 0$ | TradeOpenness > 38.0 | FXReserves > 2.0 | TotalForeignInvestment > 5.0 | ShortRate < 3.5 | FiscalSurplus > 0.2 | FinanicalDevelopment > 0.8 | Change in $Vol_{FX}$ per $Vol_{CF}$ (%): 2018 vs 2017 | No. of supporting factors (i.e. attaining the point estimate thresholds) |
|---|---|---|---|---|---|---|---|---|---|
| | Quarterly (median): 2018 | | | | | | | | |
| UK | United Kingdom | 9.8 | 0.1 | -1.5 | **0.6** | -0.7 | **0.9** | -65 | 2 |
| SG | Singapore | **55.3** | 1.3 | 1.8 | **1.3** | **0.9** | **0.8** | -46 | 4 |
| CA | Canada | 13.3 | 0.0 | 1.4 | **1.4** | -0.1 | **0.9** | -27 | 2 |
| US | United States | 5.2 | 0.0 | 0.3 | **1.8** | -1.6 | **0.9** | -26 | 2 |
| CZ | Czech Republic | 32.4 | 0.1 | 0.6 | **1.2** | **0.2** | 0.5 | -15 | 2 |
| HK | Hong Kong | **81.6** | -0.3 | **9.7** | **1.5** | **0.6** | **0.8** | -9 | 5 |
| AU | Australia | 8.6 | -0.1 | 1.2 | **1.5** | -0.2 | **0.9** | 0 | 2 |
| RU | Russia | 10.3 | 0.5 | 0.5 | 7.1 | **0.7** | 0.5 | 4 | 1 |
| KR | Korea | 16.6 | 0.2 | 1.6 | **1.5** | **0.6** | **0.8** | 7 | 3 |
| JP | Japan | 7.3 | 0.2 | 1.6 | **-0.1** | -0.6 | **0.9** | 7 | 2 |
| SW | Sweden | 16.0 | 0.0 | 0.6 | **-0.7** | 0.1 | **0.8** | 18 | 2 |
| CL | Chile | 12.1 | 0.0 | 0.1 | **2.5** | -0.4 | 0.5 | 33 | 1 |
| CH | Switzerland | 21.6 | 0.4 | 1.5 | **-0.7** | **0.4** | **1.0** | 56 | 3 |
| BR | Brazil | 5.6 | 0.1 | 0.2 | 6.4 | -1.9 | 0.6 | 57 | 0 |
| TH | Thailand | 23.7 | 0.0 | 1.2 | **1.5** | 0.0 | 0.7 | 79 | 1 |
| MA | Morocco | 15.2 | -0.1 | 0.2 | **2.3** | -0.9 | 0.4 | 103 | 1 |

*Panel B. The correlation of the number of FX resilience supporting factors and annual percentage change of FX volatility to per unit of CF volatility in 2018.*

This chart illustrates the estimated regression line between the number of supporting factors and the annual percentage change of FX volatility to per unit of CF volatility in 2018 (i.e., the far-right two columns of Panel A)

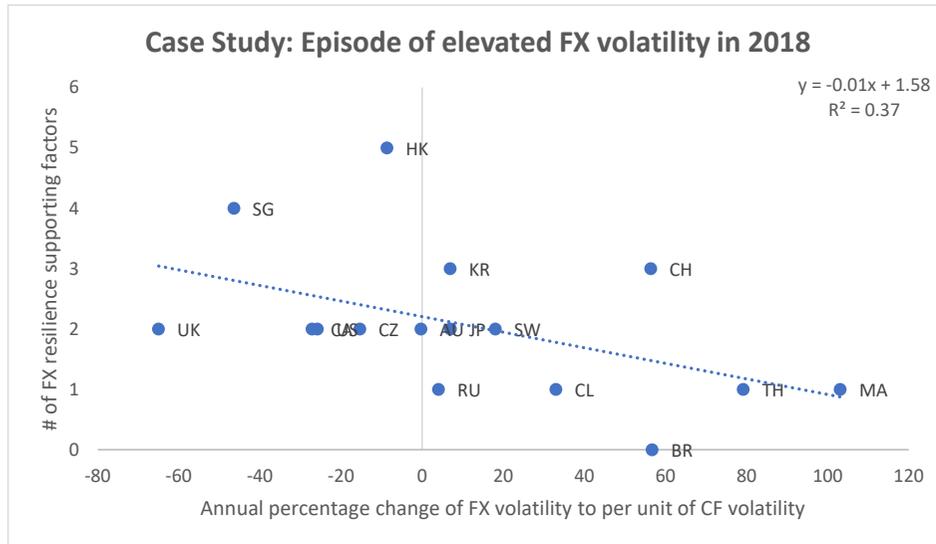

**9.**

**Case study: The 2020 Covid-19 pandemic shock**

The Covid-19 pandemic has caused volatile capital flows, in particular investment funds in 2020 (Martin *et al.*, 2020). We use this most recent global market event for a robustness test on the FX resilience measure. We proceed the case study in two steps. As with the previous case study, we compute the market-based FX resilience measure - the percentage change of FX volatility per unit of CF volatility between the 2020 average and the average of 2015-2019 for the sample countries and rank the market-based FX resilience measure from the lowest (the most resilient FX) to the highest (the least resilient FX). Then we use the median quarterly composite MF (i.e., PC1MF) in 2019[17] to estimate the FX resilience measure of the country. We rank the FX resilience measure from the lowest (the most resilient FX) to the highest (the least resilient FX), alongside the market-based FX resilience measure. Finally, we compare the ranking of the market-based FX resilience measure and the FX resilience measure, where an existence of consistency between the two rankings is encouraging to merit further application of the FX resilience measure. Table OA8 below presents the results.

Panel A of Table OA8 shows that, for EMEs, the ranking of the FX resilience measure is by and large consistent with the ranking of the market-based FX resilience measure if we consider the high level of capital control in India and China. The Czech Republic (shaded grey) is the only one exception that is ranked much higher in the theoretical FX resilience measure relative to its market-based FX resilience ranking. Similarly, Panel B of Table OA8 shows that, for AEs, both rankings are largely in line with each other, except Canada and United Kingdom (shaded grey). The small proportion of discrepancy between the two rankings for EMEs and AEs might be attributed to the different sample period used in the calculation, where the FX resilience uses the most recent available data from 2019, and the market-based FX resilience measure uses data as of 2020.

---

[17] This is the most recent data in our sample.

This case study comes with several caveats, such as excluding the element of capital control in the calculation of the market-based FX resilience measure and using different sample periods to estimate/calculate the FX resilience measure and the market-based FX resilience measure due to data availability. Nevertheless, the two measures are mostly consistent with each other in the current case study, hence providing evidence that the FX resilience measure can be a reliable measure on a country's FX resilience against capital flow shocks.

**Table OA8. Case study of FX resilience measure: the 2020 pandemic**

This table presents the FX resilience measure ranking *vs* the change in FX volatility ranking for the 20 sample countries during the 2020 pandemic. Panels A and B refer to EMEs and AEs, respectively. The FX resilience measure is estimated by using equation (7), where the values of the moderating factors are the quarterly average in 2019 (i.e., the most recent data in our sample). The change in FX volatility is calculated as the percentage change of FX volatility per unit of CF volatility between 2020 and the average of 2015 to 2019. Both the FX resilience measure and the change in FX volatility are ranked from the lowest value (i.e., the most resilient) to the highest value (i.e., the least resilient). Capital control is defined in Table OA3 in the Online Appendices. The shaded areas refer to countries with relatively large discrepancy in ranking between the FX resilience measure and the change in FX volatility.

| | FX resilience measure ranking: 2019 | | | | Market-based FX resilience ranking: 2020 |
| --- | --- | --- | --- | --- | --- |
| | $Vol_{CF}$ effect on $Vol_{FX}$ at the median PC1MF | | Capital control | | Change in $Vol_{FX}$ per $Vol_{CF}$ (%): 2020 |
| Ranking | (2019) | 90% cofidence interval | (2019) | Ranking | vs the average of [2015,2019] |
| | | *Panel A. EMEs* | | | |
| Korea | 0.00 | [-0.02, 0.01] | 0.2 | Korea | -37.49 |
| Thailand | 0.00 | [-0.01, 0.02] | 0.7 | China | -22.22 |
| Czech Republic | 0.02 | [0.00, 0.03] | 0.4 | Thailand | -19.74 |
| Malaysia | 0.02 | [0.01, 0.04] | 0.9 | India | -12.90 |
| Russia | 0.05 | [0.03, 0.06] | 0.6 | Malaysia | -7.57 |
| Chile | 0.05 | [0.03, 0.06] | 0.2 | Brazil | -4.92 |
| China | 0.05 | [0.03, 0.07] | 0.9 | Russia | 1.13 |
| Morocco | 0.06 | [0.04, 0.08] | 0.8 | Chile | 31.45 |
| Brazil | 0.07 | [0.04, 0.09] | 0.7 | CzechRepublic | 64.44 |
| India | 0.09 | [0.06, 0.11] | 1.0 | Morocco | 85.74 |
| Egypt | 0.13 | [0.09, 0.17] | 0.3 | Egypt | 95.52 |
| | | *Panel B. AEs* | | | |
| Hong Kong | -0.06 | [-0.09, -0.04] | 0.1 | Hong Kong | -54.26 |
| Singapore | -0.05 | [-0.07, -0.03] | 0.1 | UnitedKingdom | -19.06 |
| Switzerland | -0.03 | [-0.04, -0.01] | 0.3 | Sweden | -15.72 |
| Canada | 0.00 | [-0.02, 0.01] | 0.1 | Australia | -7.73 |
| Sweden | 0.00 | [-0.01, 0.02] | 0.2 | Switzerland | -2.50 |
| Australia | 0.00 | [-0.01, 0.02] | 0.3 | Singapore | -2.39 |
| Japan | 0.01 | [-0.01, 0.02] | 0.0 | United States | 10.13 |
| United Kingdom | 0.01 | [0.00, 0.03] | 0.1 | Japan | 49.13 |
| United States | 0.04 | [0.02, 0.05] | 0.1 | Canada | 85.64 |